\documentclass[a4paper]{VisionStyle}
\usepackage{epsfig}

\begin{document}

\title{Exotic sources detected in the 1~Ms Chandra Deep Field North survey}


\author{D.M.\,Alexander \and F.E.\,Bauer \and W.N.\,Brandt \and A.E.\,Hornschemeier \and C.\,Vignali \and G.P.\,Garmire \and D.P.\,Schneider}


\institute{
  Department of Astronomy \& Astrophysics, 525 Davey Laboratory, The Pennsylvania State University, University Park, PA 16802, USA }

\maketitle 

\begin{abstract}

We provide constraints on optically faint ($I\ge24$) X-ray sources, $z>6$ AGN, Very Red Objects (VROs; $I-K\ge4$), and optically faint radio sources using the 1~Ms {\it Chandra} Deep Field North survey. We argue that the majority of the optically faint X-ray sources are obscured AGN at $z=$~1--3. These sources comprise up to 50\% of the X-ray detected AGN. Approximately 30\% of the optically faint X-ray sources have properties consistent with those expected for $z>6$ AGN; however we argue that the majority of these sources lie at lower redshifts. The constraint we place on the source density of $z>6$ AGN (${\lower.5ex\hbox{$\; \buildrel < \over \sim \;$}}$~0.09~arcmin$^{-2}$) is below that predicted by some models. Approximately 30\% of the optically faint X-ray sources are also VROs. Within the $K\le20.1$ VRO population, $\approx$~15\% of the sources are AGN dominated. Conversely, we present evidence suggesting that non-AGN dominated VROs are detected at the flux limit of this 1~Ms {\it Chandra} survey. Finally, we show that $\approx$~50\% of the optically faint radio source population are detected with X-ray emission. The majority of these sources appear to be obscured AGN.

\keywords{cosmology: observations -- X-rays: background -- galaxies: AGN}

\end{abstract}

\section{Introduction}

The most recent X-ray observatories (the {\it Chandra X-ray Obsevatory}, hereafter {\it Chandra}, and {\it XMM-Newton}) are resolving close to 100\% of the $\approx$~0.5--10.0~keV background into point sources (e.g.,\ \cite{dalexander-F1:PaperV}; \cite{dalexander-F1:Hasinger01}; \cite{dalexander-F1:Giacconi02}). The majority of these X-ray sources are (in decreasing source density order) optically bright ($I<24$) Active Galactic Nuclei (AGN), luminous infrared starbursts, and ``normal'' galaxies at $z<1.5$ (e.g.,\ \cite{dalexander-F1:PaperII}; \cite{dalexander-F1:PaperXI}; A.E.~Hornschemeier et~al., in preparation); however, a significant fraction ($\approx35$\%) have optically faint (i.e.,\ $I\ge24$) counterparts (e.g.,\ \cite{dalexander-F1:PaperVI}). These optically faint X-ray sources are too faint for practical optical spectroscopy, although their properties suggest that many are luminous AGN at $z>1$ (e.g.,\ \cite{dalexander-F1:Fabian00}; \cite{dalexander-F1:PaperVI}; \cite{dalexander-F1:Barger01a}; \cite{dalexander-F1:Cowie01}; \cite{dalexander-F1:Gandhi02}; \cite{dalexander-F1:Koekemoer02}). Thus, the X-ray emission from these sources gives us a window on the accretion power of the Universe at moderate-to-high redshift. However, the properties of a fraction of these sources are also consistent with those expected for $z>6$ AGN (\cite{dalexander-F1:PaperVI}) and non-AGN dominated sources at $z\approx$~1 (\cite{dalexander-F1:PaperX}). These sources are important for constraining the space density of the first super-massive black holes (SMBHs) and determining the role of AGN in galaxy formation and evolution. As very few examples of these source types had been found prior to {\it Chandra} and {\it XMM-Newton} surveys, we collectively refer to them here as exotic sources.

In this contribution we present constraints on exotic sources detected in a 70.3~arcmin$^{2}$ region (hereafter referred to as the reduced Hawaii flanking field area) of the 1~Ms {\it Chandra} Deep Field North (CDF-N; \cite{dalexander-F1:PaperV}) survey. The $\approx$~5.3~arcmin$^{2}$ Hubble Deep Field North (HDF-N) region lies at the aim-point of the CDF-N survey and has some of the deepest multi-wavelength observations and optical spectroscopy in the entire sky (e.g.,\ \cite{dalexander-F1:Ferguson00}). In \S\ref{dalexander-F1:obs} we briefly describe the CDF-N observations and X-ray-to-optical source matching. In \S\ref{dalexander-F1:optfaint} we provide constraints on the properties of optically faint X-ray sources and argue that the majority are obscured AGN at $z=$~1--3. In \S\ref{dalexander-F1:zgtsix} we provide constraints on the number of $z>6$ AGN and give evidence that there are fewer sources than predicted by the \cite*{dalexander-F1:Haiman99} model. In \S\ref{dalexander-F1:vros} we provide constraints on the X-ray properties of Very Red Objects (VROs; $I-K\ge4$) and determine the AGN fraction within the $K\le20.1$ VRO population. In \S\ref{dalexander-F1:radiosubmm} we provide constraints on the X-ray properties of optically faint radio sources and show that up to 50\% are likely to host obscured AGN. Finally, in \S\ref{dalexander-F1:discussion} we discuss our results and planned deeper observations. The results presented in this contribution are taken primarily from \cite*{dalexander-F1:PaperVI} and \cite*{dalexander-F1:PaperX}.

\section{Observations and source matching}
\label{dalexander-F1:obs}

\begin{figure*}[t]
  \begin{center}
    \epsfig{file=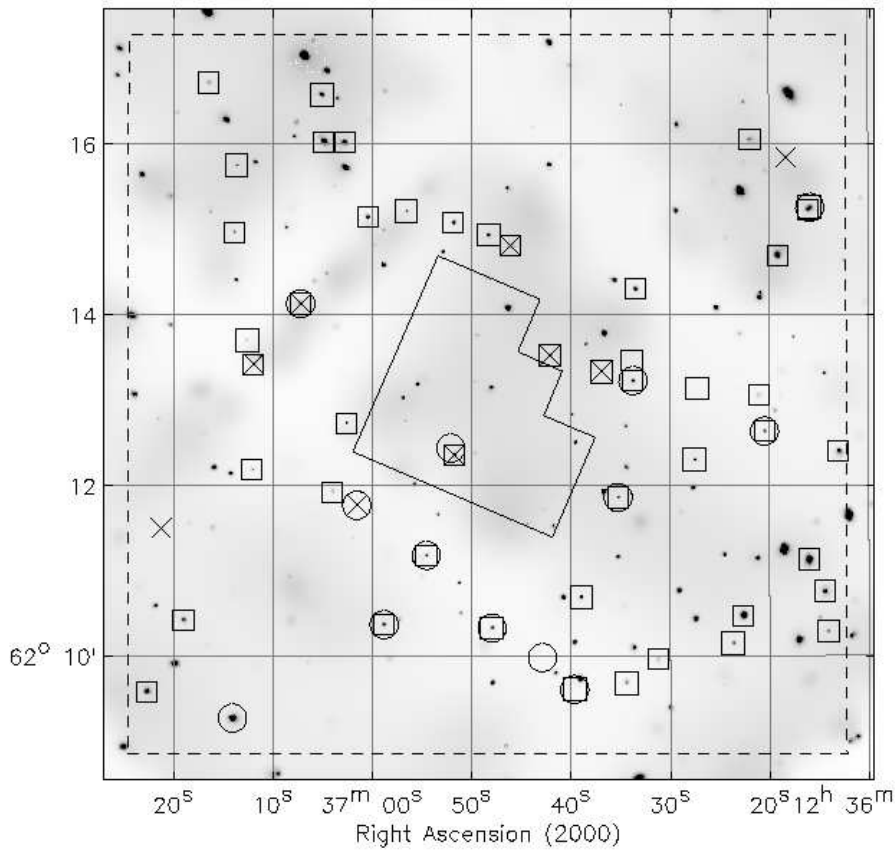,width=12.0cm}
  \end{center}
\caption{\small{Adaptively smoothed full-band {\it Chandra} image. The squares show the positions of the 47 optically faint X-ray sources, the circles show the positions of the 13 X-ray detected VROs, and the crosses show the positions of the 9 X-ray detected optically faint radio sources. This image has been adaptively smoothed at the $3\sigma$ level using the code of Ebeling, White, \& Rangarajan (2002). The HDF-N is shown as the polygon at the center of the image, and the dashed box indicates the 70.3 arcmin$^2$ reduced Hawaii flanking field area. Most of the apparent diffuse emission is instrumental background. The faintest X-ray sources are below the significance level of the smoothing and thus are not visible in this figure.}}
\label{dalexander-F1_fig:fig1}
\end{figure*}

The X-ray observations used in these studies were obtained with the {\it Chandra} Advanced CCD Imaging Spectrometer (ACIS; G.P. Garmire et~al., in preparation) on-board {\it Chandra}. With the exception of a number of lower significance X-ray sources reported in \S\ref{dalexander-F1:vros} and \S\ref{dalexander-F1:radiosubmm}, all of the X-ray sources were taken from \cite*{dalexander-F1:PaperV}.

In total 141 sources (hereafter referred to as the entire X-ray sample) are detected in the reduced Hawaii flanking field area down to 0.5--2.0~keV (soft-band) and 2--8~keV (hard-band) flux limits of \hbox{$\approx 3\times 10^{-17}$~erg~cm$^{-2}$~s$^{-1}$} and \hbox{$\approx 2\times 10^{-16}$~erg~cm$^{-2}$~s$^{-1}$} at the aim point [see Figure~\ref{dalexander-F1_fig:fig1} for the 0.5--8.0~keV (full-band) {\it Chandra} image]. The absolute X-ray source positions within 5~arcmin of the aim point are accurate to better than 0.6~arcsec; for sources outside this region, the positional errors rise to better than $\approx 1$~arcsec (see \cite{dalexander-F1:PaperV}).

The optical and near-IR catalogs used in this study were constructed from the $I$-band and $HK^{\prime}$-band images of \cite*{dalexander-F1:Barger99}. These images reach $\approx 2\sigma$ magnitude limits of $I=25.3$ and $HK^{\prime}=21.4$, respectively.\footnote{The relationship between the $K$-band and $HK^\prime$-band is $HK^\prime-K=0.13+0.05(I-K)$ (\cite{dalexander-F1:Barger99}).} The X-ray sources were matched to $I$-band counterparts using a search radius of 1~arcsec. Forty-seven X-ray sources are optically faint (i.e.,\ $I\ge24$), including 15 undetected down to $I=25.3$ (see Figure~\ref{dalexander-F1_fig:fig1} and Figure~\ref{dalexander-F1_fig:fig2}).

\section{Optically faint X-ray sources}
\label{dalexander-F1:optfaint}

\begin{figure*}[t]
  \begin{center}
    \epsfig{file=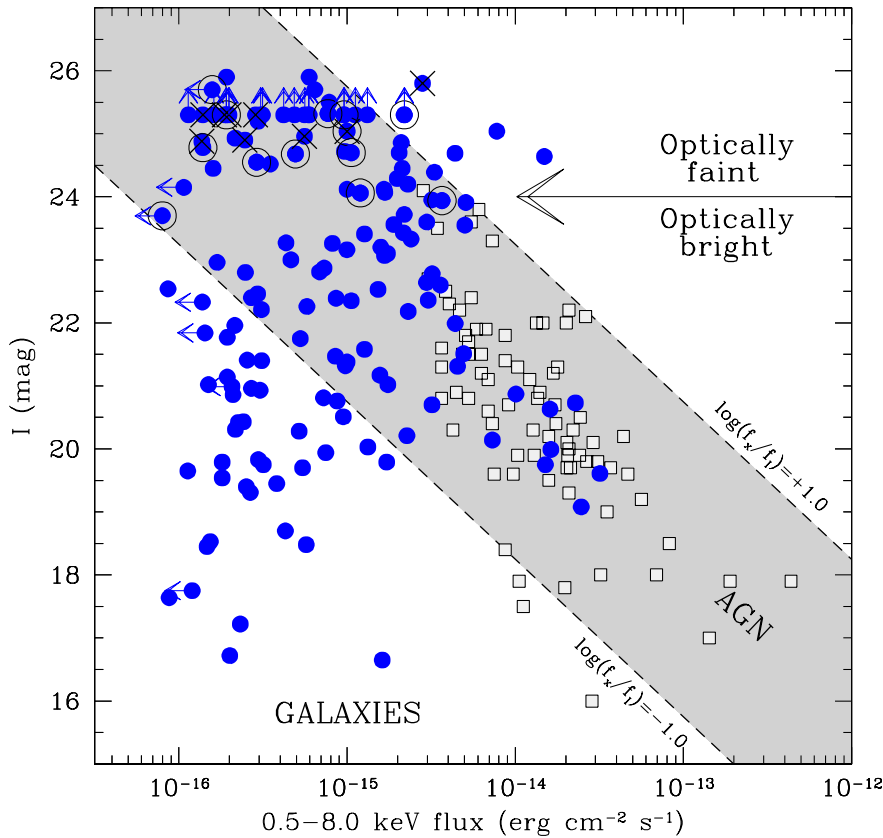, width=12cm}
  \end{center}
\caption{\small{$I$-band magnitude versus full-band flux for sources detected in the {\it ROSAT} Ultra-Deep Survey of the Lockman Hole (squares; Lehmann et~al. 2001) and in the reduced Hawaii flanking field area of the CDF-N survey (filled circles; Alexander et~al. 2001). The {\it ROSAT} 0.5--2.0~keV fluxes were converted to 0.5--8.0~keV assuming $\Gamma=2.0$, and the $R$-band magnitudes were converted to $I$-band assuming $R-I=0.9$ (see \S4.2 of Alexander et~al. 2001 for justification). The 15 X-ray sources without optical counterparts down to $I=25.3$ are shown as upper limits; these sources are potential $z>6$ AGN (see \S\ref{dalexander-F1:zgtsix}). The 13 X-ray detected VROs are indicated by large open circles (see \S\ref{dalexander-F1:vros}). The 9 X-ray detected optically faint radio sources are indicated by crosses (see \S\ref{dalexander-F1:radiosubmm}). The diagonal lines indicate constant flux ratios. The approximate range in flux ratios for typical AGN and galaxies is shown.}}
\label{dalexander-F1_fig:fig2}
\end{figure*}

\begin{figure}[t]
  \begin{center}
    \epsfig{file=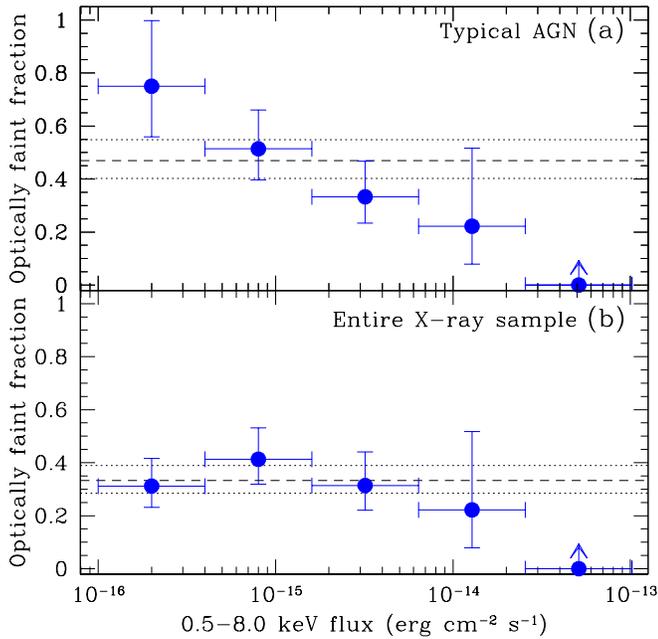, width=9cm}
  \end{center}
\caption{\small{The fraction of optically faint X-ray sources versus full-band flux for (a) typical AGN and (b) the entire X-ray sample. The filled circles show the optically faint X-ray source fraction, the width of each X-ray flux bin is shown by bars in the $x$-axis direction and the 1$\sigma$ uncertainty in the source fraction is shown by bars in the $y$-axis direction. The dashed line shows the overall fraction of optically faint X-ray sources and the dotted lines show the 1$\sigma$ uncertainty on this value.}}
\label{dalexander-F1_fig:fig3}
\end{figure}

The optically faint X-ray sources account for $33^{+6}_{-5}$\% of the entire X-ray sample. The fraction is constant at $\approx 35$\% for full-band fluxes $<10^{-14}$~erg~cm$^{-2}$~s$^{-1}$ (see Figure~\ref{dalexander-F1_fig:fig3}). This counter-intuitive result is due to the emergence of a population of luminous infrared starbursts and ``normal'' galaxies at faint X-ray fluxes (e.g.,\ \cite{dalexander-F1:PaperXI}; A.E. Hornschemeier et~al, in preparation), and the optically faint X-ray source fraction would rise at faint X-ray fluxes if these sources were absent (see Figure~\ref{dalexander-F1_fig:fig3}a). The constraints at full-band fluxes $>10^{-14}$~erg~cm$^{-2}$~s$^{-1}$ are poor due to the small areal coverage of this survey. Wide-field, shallow X-ray surveys such as ChaMP (\cite{dalexander-F1:Wilkes01}) and the {\it XMM-Newton} serendipitous surveys (\cite{dalexander-F1:Watson01}; \cite{dalexander-F1:Baldi02}) are better suited to determining the fraction of optically faint sources at brighter X-ray fluxes. This important constraint could have an impact on the optical identification of X-ray sources detected in surveys prior to {\it Chandra} and {\it XMM} (e.g.,\ {\it ASCA} and {\it BeppoSAX} surveys), where the typical positional uncertainty of an X-ray source can produce up to 50 possible optical counterparts at $I=24$.

All of the optically faint X-ray sources have X-ray-to-optical flux ratios consistent with or greater than those typically found for typical AGN [i.e.,\ $-1<\log{({{f_{\rm X}}\over{f_{\rm I}}})}<+1$; see Figure~\ref{dalexander-F1_fig:fig2}]. This evidence suggests that the majority (possibly all) of the optically faint X-ray sources are AGN. Indeed, direct evidence for AGN activity is found from X-ray variability measurements for $\approx$~40\% of the X-ray brightest optically faint sources. As 53 optically bright X-ray sources have X-ray-to-optical flux ratios consistent with those found for typical AGN, $\approx$~50\% of the X-ray detected AGN could be optically faint (see Figure~\ref{dalexander-F1_fig:fig3}a). This fraction should be taken as an upper limit as some optically bright AGN will have $\log{({{f_{\rm X}}\over{f_{\rm I}}})}<-1$ and a fraction of the optically faint X-ray source population may not be AGN (e.g.,\ see \S\ref{dalexander-F1:vros}).

\begin{figure}[t]
  \begin{center}
    \epsfig{file=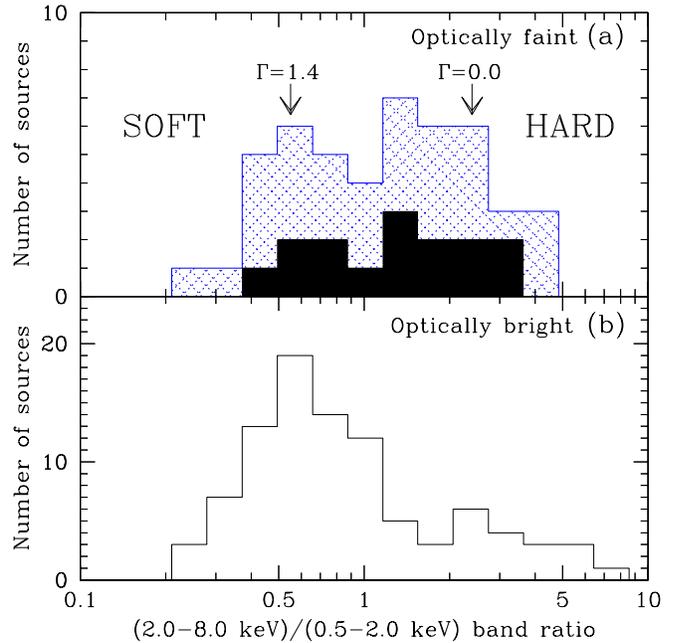, width=9cm}
  \end{center}
\caption{\small{X-ray band ratio distributions of (a) the optically faint sample and (b) the optically bright sample. Clearly a larger fraction of optically faint X-ray sources have flat X-ray spectral slopes. The two X-ray band ratio distributions are distinguishable according to the Kolmogorov-Smirnov test at the 99.4\% significance level. The solid histogram in the top panel shows the overlaid X-ray band ratio distribution of the 15 sources without optical counterparts down to $I=25.3$.}}  
\label{dalexander-F1_fig:fig4}
\end{figure}

In the Unified AGN model (e.g.,\ \cite{dalexander-F1:Antonucci93}), AGN are simplistically divided into obscured and unobscured sources. In terms of this model, the average effective photon index of the optically faint X-ray sources ($\Gamma\approx0.9$) suggests the majority are obscured AGN (see Figure~\ref{dalexander-F1_fig:fig4}). Assuming the underlying emission is typical of an unobscured AGN (e.g.,\ $\Gamma=2.0$; \cite{dalexander-F1:George00}), the average X-ray band ratio (the ratio of the hard-band to soft-band count rate) of the optically faint X-ray sources corresponds to an intrinsic absorption column density of \hbox{$N_{\rm H}\approx 1.5\times10^{23}$ cm$^{-2}$} at $z=2$. Direct evidence for such heavy X-ray absorption is found for some optically faint sources bright enough for X-ray spectral analysis (e.g.,\ \cite{dalexander-F1:PaperIV}; \cite{dalexander-F1:Crawford02}). These results suggest that the majority of the optically faint X-ray source population are obscured AGN. 

\begin{figure}[t]
  \begin{center}
    \epsfig{file=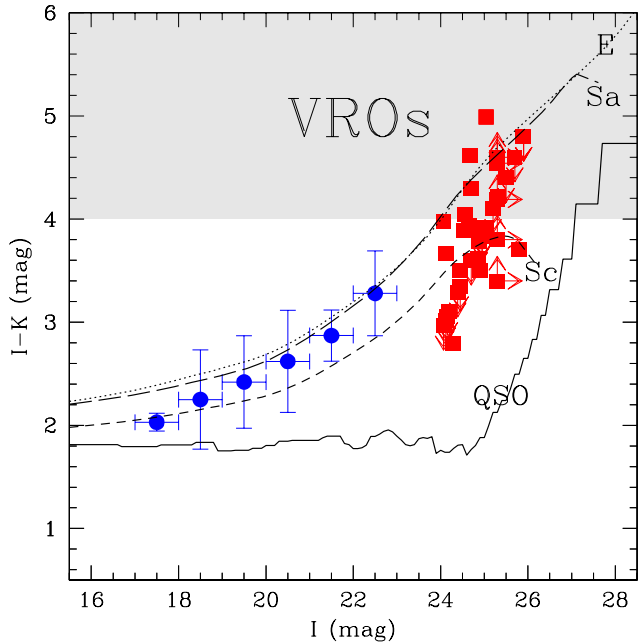, width=9cm}
  \end{center}
\caption{\small{$I-K$ colour versus $I$-band magnitude. The filled squares are the optically faint X-ray sources. The filled circles are the average $I-K$ colours for the $I<23$ X-ray sources. The dotted, long-dashed, and short-dashed curves are the redshift tracks of $M_{I}^{*}$ E, Sa, and Sc host galaxies, respectively. The solid curve is the redshift track of an $M_{I}=-23$ QSO. At optically faint magnitudes $\approx$~30\% of the X-ray source population with measurable constraints are VROs. See Alexander et~al. (2001, 2002b) for further details.}}  
\label{dalexander-F1_fig:fig5}
\end{figure}

The redder $I-K$ colours of the optically faint X-ray sources are consistent with the positive $K$-corrections of a normal galaxy at $z>1$, providing evidence that the optically faint X-ray sources are the continuation of the optically bright X-ray source population to redshifts of $z=$~1--3 (see Figure~\ref{dalexander-F1_fig:fig5}; see also \cite{dalexander-F1:Cowie01}).\footnote{In obscured AGN the optical emission is usually dominated by the host galaxy.} This is in good agreement with the redshift range found for the six optically faint X-ray sources with redshift determinations (i.e.,\ $z=$~0.9--4.4; see Figure~\ref{dalexander-F1_fig:fig6}).\footnote{Three sources have spectroscopic redshifts, and three sources have photometric or millimetric redshifts; see \cite*{dalexander-F1:PaperVI} and \cite*{dalexander-F1:PaperX}.} 

There is considerable interest in the space density of heavily obscured luminous AGN (i.e., obscured QSOs), since such sources are expected by the Unified AGN model but few are found locally (e.g.,\ \cite{dalexander-F1:Halpern99}; \cite{dalexander-F1:Franceschini00}; and references therein). Many X-ray background synthesis models predict a large number of obscured QSOs (e.g.,\ $L_X>3\times10^{44}$~erg s$^{-1}$) at high redshift (e.g.,\ \cite{dalexander-F1:Wilman00}; \cite{dalexander-F1:Gilli01}), although very few sources have been found in the optically bright X-ray source population; however, see \cite*{dalexander-F1:Norman02}. Therefore, if these sources exist, the majority are likely to be optically faint X-ray sources. Only one of our six optically faint X-ray sources with redshifts is likely to be an obscured QSO. This $I=25.8$ source lies in the HDF-N itself and has a photometric redshift of $z=2.75^{+0.13}_{-0.20}$ (90\% confidence level) and an unabsorbed rest-frame full-band luminosity of $3\times10^{44}$ erg s$^{-1}$ (see \cite{dalexander-F1:Fabian00}; \cite{dalexander-F1:Cowie01}; \cite{dalexander-F1:Crawford02}; \cite{dalexander-F1:Gandhi02}; \cite{dalexander-F1:Stern02} for some other optically faint obscured QSO candidates). Although we do not have redshift constraints for the majority of our optically faint X-ray sources, based on luminosity agruments, $\approx$~5--20\% could be obscured QSOs assuming an average redshift of $z=2$. The upper limit determination assumes the X-ray absorption is Compton thick (i.e.,\ $N_H>1.5\times10^{24}$ cm$^{-2}$) and that the observed X-ray emission is scattered; see also \cite*{dalexander-F1:Fabian02} for predictions of the number of Compton thick QSOs in a 1~Ms {\it Chandra} survey.

\section{AGN at $z>6$}
\label{dalexander-F1:zgtsix}

\begin{figure}[t]
  \begin{center}
    \epsfig{file=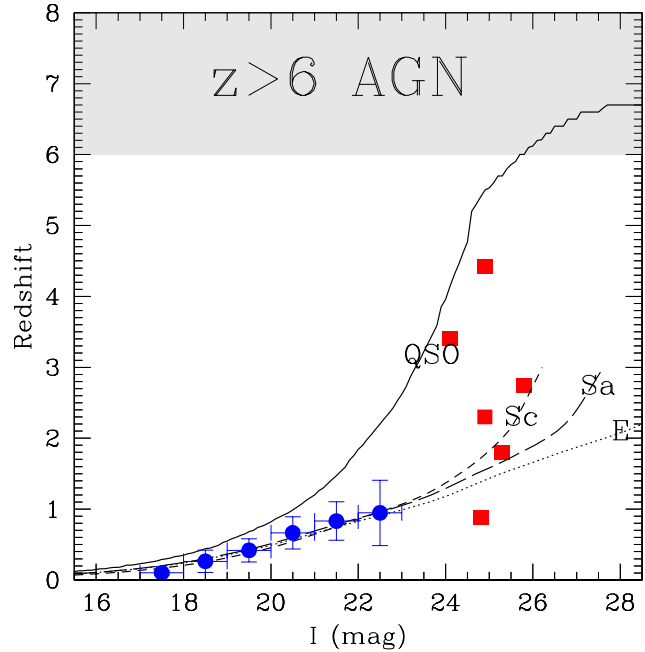, width=9cm}
  \end{center}
\caption{\small{X-ray source redshifts versus $I$-band magnitude. The filled squares are the optically faint X-ray sources with spectroscopic, photometric or millimetric redshifts. The filled circles are the average spectroscopic redshifts for the $I<23$ X-ray sources. The curves have the same meaning as in Figure~\ref{dalexander-F1_fig:fig5}. A $z>6$ AGN would be essentially undetected in the $I$-band as the Lyman-$\alpha$ forest and Gunn-Peterson trough will absorb essentially all the optical flux. See Alexander et~al (2001, 2002b) for further details.}}
\label{dalexander-F1_fig:fig6}
\end{figure}

One of the major goals of observational cosmology is to constrain the mass and space density of the first SMBHs and determine their role in galaxy formation and evolution. The probable redshift range of these first objects is $z=$~6--15, and such sources are considered to be the first AGN (\cite{dalexander-F1:Haiman99}). At $z>6$ the Lyman-$\alpha$ forest and Gunn-Peterson trough will absorb essentially all flux through the $I$ band and an AGN would appear optically blank (e.g.,\ see Figure~\ref{dalexander-F1_fig:fig6}). However, a $z>6$ AGN would be visible at X-ray energies due to the accretion of material around the SMBH; at the flux limit of this 1~Ms {\it Chandra} survey, a $z>6$ AGN will have a rest-frame full-band luminosity of $>5\times10^{43}$ erg s$^{-1}$.

Fifteen X-ray sources do not have optical counterparts down to $I=25.3$ (see Figure~\ref{dalexander-F1_fig:fig2}). This number of optically blank X-ray sources is consistent with the expected number of $z>6$ AGN predicted by \cite*{dalexander-F1:Haiman99} on the basis of a hierarchical cold-dark matter model. However, in \cite*{dalexander-F1:PaperVI} we argued that the majority of these optically blank X-ray sources are the extension of the optically faint X-ray source population to fainter optical magnitudes; we briefly discuss the the evidence here.

\begin{figure*}[!t]
  \begin{center}
    \epsfig{file=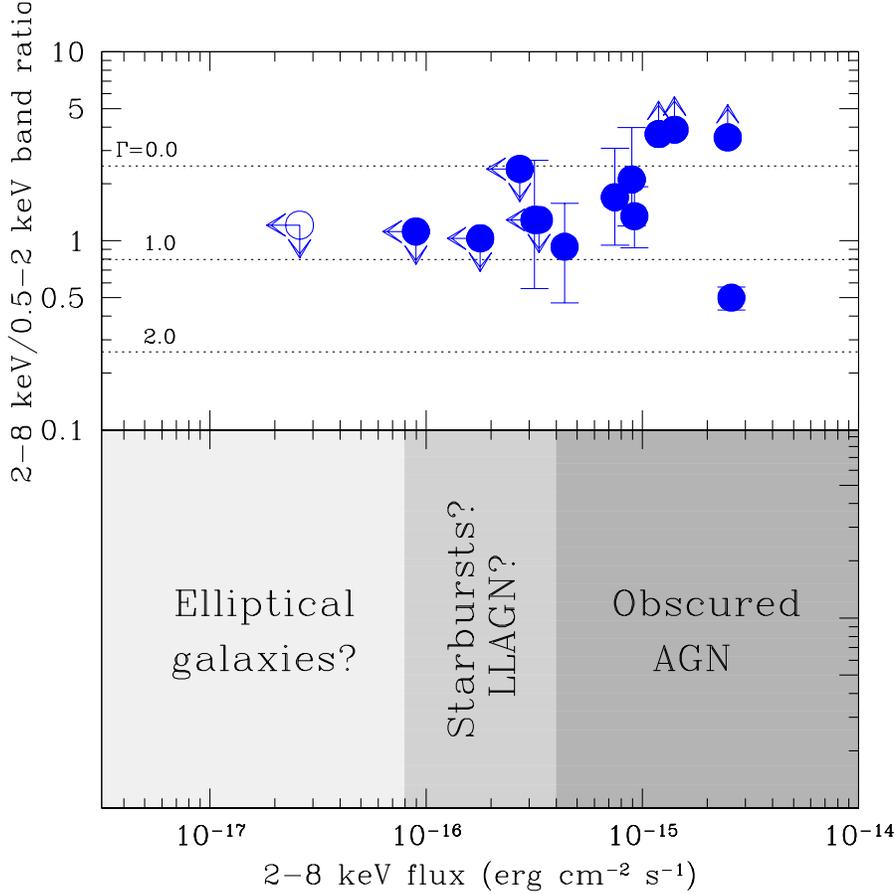, width=12cm}
  \end{center}
\caption{\small{Hard-band flux versus the X-ray band ratio for the VROs. The filled circles indicate the X-ray detected VROs. The open circle indicates the stacking-analysis results for the X-ray undetected sources from the $K\le20.1$ VRO sample. The equivalent photon indices ($\Gamma$) are shown on the left-hand side of the top panel. The bottom panel shows the approximate range of hard-band fluxes for the different VRO source types detected; see Alexander et~al. (2002b) for further details.}}  
\label{dalexander-F1_fig:fig7}
\end{figure*}

The $\approx$~5.3~arcmin$^{2}$ HDF-N region has considerably deeper optical coverage than the rest of the reduced Hawaii flanking field area. Within the HDF-N itself we detected an $I=25.8$ AGN at $z=2.75$ (see \S\ref{dalexander-F1:optfaint}) that would have been undetected, and thus considered optically blank, if it lay outside this region.\footnote{We also note that we have detected an $I\approx$~27 source in the current 1.38~Ms {\it Chandra} observation (W.N.~Brandt et~al., in preparation).} Although the statistics are limited, based on the area of the HDF-N itself, we would expect $\approx 13$ such sources within the reduced Hawaii flanking field area. This result suggests that almost all of the optically blank X-ray sources are optically faint X-ray sources at fainter optical magnitudes and thus lie at $z\approx$~1--3. Further evidence for this is found from  Kolmogorov-Smirnov tests of the X-ray band ratio and full-band flux distributions. These tests show that the optically blank and optically faint X-ray source distributions are indistinguishable with 83\% and 7\% confidence, respectively. The high probability found for the X-ray band ratio distributions suggest that both source populations contain the same object types (i.e.,\ mostly obscured AGN), while the low probability for the full-band flux distribution is due to the fainter X-ray fluxes of the optically blank X-ray sources. Indeed, assuming typical X-ray-to-optical flux ratios, an AGN at the flux limit of this 1~Ms {\it Chandra} survey could have $I=28.5$. From this analysis we argue that there are ${\lower.5ex\hbox{$\; \buildrel < \over \sim \;$}}$~6 AGN at $z>6$ in the reduced Hawaii flanking field area, corresponding to a source density of ${\lower.5ex\hbox{$\; \buildrel < \over \sim \;$}}$~0.09~arcmin$^{-2}$. Although inconsistent with the model of \cite*{dalexander-F1:Haiman99}, this source density is not inconsistent with the Model D result of \cite*{dalexander-F1:Gilli01} which predicts $<1$ AGN at $z>6$ in this region.

\section{Very Red Objects}
\label{dalexander-F1:vros}

A large fraction of the optically faint X-ray source population has red colours, and $\approx$~30\% of the optically faint X-ray sources with measurable colours are VROs (see Figure~\ref{dalexander-F1_fig:fig5}).\footnote{In many studies these sources are also referred to as Extremely Red Objects (EROs).} Current constraints suggest the VRO population is comprised of passively evolving elliptical galaxies and dusty starbursts at $z\approx$~1 (e.g.,\ \cite{dalexander-F1:Cimatti98}; \cite{dalexander-F1:Moriondo00}), although few constraints exist on the fraction of VROs with AGN activity (see \cite{dalexander-F1:PaperII} for the best AGN constraints prior to this study). In \cite*{dalexander-F1:PaperX} we defined complete VRO samples to determine the AGN fraction in the VRO population and to further constrain the nature of optically faint X-ray sources.

Thirteen VROs are detected with X-ray emission (see Figure~\ref{dalexander-F1_fig:fig2}): nine are high-significance X-ray sources from \cite*{dalexander-F1:PaperV}, and the other four sources were matched to lower significance X-ray sources. From this result the fraction of the $K\le20.1$ VRO population with X-ray emission is $21^{+12}_{-8}$\%. Since almost all of these sources are optically faint, the majority have X-ray properties consistent with obscured AGN activity. Indeed, the average effective photon index ($\Gamma\approx$~0.9) of these sources suggest that they are the ``Very Red'' tail of the optically faint X-ray source population. However, the four VROs detected close to the X-ray flux limit have different X-ray and optical properties. None of these four sources is detected in the hard band and, with a stacked-average effective photon index of $\Gamma>$~1.4, they are clearly different from the other X-ray detected VROs (see Figure~\ref{dalexander-F1_fig:fig7}). These sources also tend to have irregular optical morphologies and extremely red colours (i.e.,\ $I-K>5$). Better constraints can be placed on two of the sources for which we have redshifts and/or multi-wavelength data. One source lies at $z=0.884$ and has a spectral energy distribution consistent with that found for Arp~220, the archetypal dusty starburst. However, we are unable to determine whether the X-ray emission from this VRO is produced by star formation activity or low-luminosity AGN (LLAGN) activity; two further VROs have similar optical and X-ray properties to this source. The other VRO resides in the HDF-N region and has a photometric redshift of $z=1.0$. The optical morphology, optical-to-near-IR spectral energy distribution, and X-ray properties of this source are consistent with those expected for a $z=1.0$ elliptical galaxy. Thus, these results suggest we have reached the X-ray fluxes necessary to detect non-AGN dominated VROs. Since the majority of these non-AGN dominated VROs have $I\ge24$, this also shows that a fraction of the optically faint X-ray source population are not AGN. From these results, the fraction of the $K\le20.1$ VRO population with AGN activity is $14^{+11}_{-7}$\%.

The majority of the VROs are undetected in the X-ray band. However, using stacking analysis techniques (see \cite{dalexander-F1:PaperIV} for details), we found a statistical detection in the soft and full bands for the X-ray undetected VROs (see Figure~\ref{dalexander-F1_fig:fig7}). The properties of this average X-ray emission are similar to those found for the X-ray detected $z=1.0$ elliptical galaxy and are consistent with those expected for $M^{*}_{I}$ elliptical galaxies; see also \cite*{dalexander-F1:PaperII}. An $\approx$~5~Ms {\it Chandra} exposure is required to detect the individual X-ray emission from a large number of these sources.

\section{Optically faint radio sources}
\label{dalexander-F1:radiosubmm}

Approximately 20\% of the radio sources within this region are optically faint (\cite{dalexander-F1:Richards99}). \cite*{dalexander-F1:Richards99} proposed that the optically faint radio source population could be composed of: (1) luminous dust-enshrouded starbursts at $z\approx$~1--3, (2) luminous obscured AGN at $z{\lower.5ex\hbox{$\; \buildrel > \over \sim \;$}}$~2, or (3) AGN at $z>6$. Sub-mm (850~$\mu$m) observations have shown that a large fraction ($\approx 50$\%) of the optically faint radio source population appears to host dusty starburst activity at $z\approx$~1--3 (e.g.,\ \cite{dalexander-F1:Barger00}; \cite{dalexander-F1:Chapman01}). However, in \cite*{dalexander-F1:PaperVI} we showed that nine ($53^{+24}_{-17}$\%) of the 17 optically faint radio sources reported in \cite*{dalexander-F1:Richards99} have X-ray emission (see Figure~\ref{dalexander-F1_fig:fig2}): six are high-significance X-ray sources from \cite*{dalexander-F1:PaperV}, and the other three sources were matched to lower significance X-ray sources. The majority of these sources appear to be obscured AGN, showing that models of the optically faint radio source population should take into account the contributions from both AGN and starburst activity.

Are any of the X-ray sources also sub-mm sources or are these two distinct populations? In \cite*{dalexander-F1:PaperVI} we found that two of the X-ray detected optically faint radio sources also had sub-mm emission. Interestingly, these two sources are the faintest X-ray sources, and their soft X-ray emission and moderate X-ray luminosities are not inconsistent with luminous starburst activity. However, a more extensive study of the sub-mm properties of the X-ray sources in this region uncovered a larger number of X-ray detected optically faint radio sources with sub-mm emission (\cite{dalexander-F1:Barger01b}). The X-ray properties of these sources are consistent with obscured AGN activity, suggesting that the majority of the X-ray detected sub-mm sources host AGN. While these results place constraints on the nature of the X-ray emission from sub-mm sources, the origin of the sub-mm emission remains unknown.

Finally, we note the detection of an extended X-ray source with a ``disturbed'' X-ray morphology but without any obvious optically bright counterparts (see \cite{dalexander-F1:PaperIX}). This source does, however, have an overdensity of exotic sources, including optically faint radio and X-ray sources, VROs, and sub-mm sources. Although the current constraints are weak, this source may be a moderate-to-high redshift cluster (i.e.,\ $z{\lower.5ex\hbox{$\; \buildrel > \over \sim \;$}}$~1). This tentative link between clustering needs to be further investigated.

\section{Discussion and future observations}
\label{dalexander-F1:discussion}

The properties of the majority of the optically faint X-ray sources detected in the 1~Ms CDF-N survey are consistent with those expected from obscured AGN activity at $z=$~1--3. These sources are likely to be the continuation of the optically bright X-ray source population to fainter $I$-band magnitudes. Fifteen of these sources have optical properties consistent with those expected for $z>6$ AGN. However, almost all are likely to lie at lower redshifts, placing a constraint on the source density of $z>6$ AGN of ${\lower.5ex\hbox{$\; \buildrel < \over \sim \;$}}$~0.09~arcmin$^{-2}$. Approximately 30\% of the optically faint X-ray sources with measurable colours are VROs, and the fraction of the $K\le20.1$ VRO population with AGN activity is $14^{+11}_{-7}$\%. These AGN-dominated VROs appear to be the ``Very Red'' tail of the optically faint X-ray source population. Conversely, four VROs detected close to the X-ray flux limit have X-ray properties consistent with non-AGN dominated sources. Approximately 50\% of the optically faint radio sources are detected with X-ray emission and the majority of these sources appear to be obscured AGN. Some of these sources also have sub-mm counterparts. Examples of all of these exotic source types have been detected near a potential moderate-to-high redshift cluster (see \cite{dalexander-F1:PaperIX}).

The number of optically bright X-ray sources with X-ray-to-optical flux ratios consistent with AGN activity is similar to the number of optically faint X-ray sources, suggesting that  up to 50\% of the X-ray detected AGN are optically faint. Since the majority of the optically bright X-ray source population lies at $z<1.5$, the bulk of the moderate-to-high redshift accretion activity is probably produced by optically faint X-ray sources. Thus, determining the redshifts of the optically faint X-ray sources is vital in order to constrain the moderate-to-high redshift accretion history of the Universe. Due to their extreme optical faintness, the photometric redshift technique is the most efficient method of redshift determination for these sources (e.g.,\ \cite{dalexander-F1:Fernandez99}). The Great Observatories Origins Deep Survey (GOODS) project will soon obtain deep, public {\it HST} Advanced Camera for Surveys and {\it SIRTF} coverage over $\approx$~160 arcmin$^2$ of the most sensitive region of the CDF-N.\footnote{See http://www.stsci.edu/science/goods/ for details.} These observations will allow photometric redshift determinations for the majority of the optically faint X-ray source population and will place significantly tighter constraints on the space density of $z>6$ AGN.

The CDF-N survey will be extended to $\approx$~2~Ms in February 2002. Since {\it Chandra} observations are not background limited at $\approx$~2~Ms, this exposure will be two times more sensitive than the observations presented in this contribution. This deeper exposure should detect more examples of optically faint non-AGN dominated sources and will be more sensitive to higher redshift optically faint X-ray sources and lower luminosity $z>6$ AGN.

\begin{acknowledgements}

This work would not have been possible without the support of the entire {\it Chandra} and ACIS teams. We acknowledge the financial support of NASA grant NAS~8-38252 (GPG, PI), NSF CAREER award AST-9983783 (DMA, FEB, WNB, CV), NASA GSRP grant NGT5-50247 (AEH), and NSF grant AST-9900703~(DPS).

\end{acknowledgements}

\end{document}